\documentclass[aps,twocolumn,pra,superscriptaddress,letterpaper]{revtex4-1}
\usepackage{amssymb}
\usepackage{mathtools}
\usepackage[pdftex]{graphicx}
\usepackage[usenames, dvipsnames, svgnames, table]{xcolor}
\usepackage[colorlinks=true, citecolor=Blue,
            linkcolor=BrickRed, urlcolor=ForestGreen]{hyperref}
\usepackage{txfonts}
\usepackage{mathrsfs}
\usepackage{bm}
\usepackage{multirow}
\usepackage{color}
\usepackage{comment}

\begin{document}


\title{Verification of the Born rule via direct measurement of superposition wavefunction}

\author{Meng-Jun Hu}
\email{humj@baqis.ac.cn}
\affiliation{Beijing Academy of Quantum Information Sciences, Beijing 100193, China}

\date{\today}

\begin{abstract}
The Born rule, which is one of foundational axioms of quantum theory, states that the probability of obtain outcome $a$ for the quantum state $|\psi\rangle$ is determined by $P(a)=|\langle a|\psi\rangle|^{2}$. Despite its great success in predicting the experimental outcomes, there still lacks a direct way to verify the Born rule. Here, we show that the weak value based direct measurement of superposition wavefunction is feasible, which can be used to verify the Born rule directly. The plausible experiment is suggested.
\end{abstract}

\maketitle


\section{Introduction}
For a quantum system described by the state $|\psi\rangle$, if we perform the measurement of observable $\hat{A}$, the Born rule tells us that the probability of obtain outcome $a$ is given by $P(a)=|\langle a|\psi\rangle|^{2}$. The Born rule has already withstood numerical experimental tests and there seems no question for its correctness. The interesting and important question now is whether or not the Born rule can be directly verified. This is possible if we can directly measure the probability amplitude $\langle a|\psi\rangle$. For the observable to be position $\hat{X}$, it implies the direct measurement of wavefunction $\psi(x)$. Since the quantum state can be superposed, the best way to verify the Born rule is to realize  direct measurement of superposition wavefunction $\psi(x_{1})+\psi(x_{2})$ and $\psi(x_{1}), \psi(x_{2})$ respectively, and then check if they satisfy the relation 
\begin{equation}\nonumber
|\psi(x_{1})+\psi(x_{2})|^{2}=|\psi(x_{1})|^{2}+|\psi(x_{2})|^{2}+\psi(x_{1})\psi^{*}(x_{2})+\psi(x_{2})\psi^{*}(x_{1}).    
\end{equation}

The possibility of direct measurement of wavefunction is attributed to the introduction of weak value by Aharonov, Albert and Vaidman (AAV) in 1988 \cite{AAV}. Lundeen {\it et al} first demonstrated experimentally the direct spatial wavefunction reconstruction of photons via weak value in 2011 \cite{Lundeen}, and serious works on this topic has been reported since then \cite{wave1, wave2, wave3, wave4}.
In the original work of AAV, the weak value is defined based on the framework of weak measurements. It has been realized recently that the weak value is independent of measurement strength, which not only extent the physical meaning of weak value, but also makes the practical measurement more accurate \cite{wave5}. To date, however, direct measurement of superposition wavefunction has not been reported. 

Here we show that the direct measurement of superposition wavefunction based on weak value is feasible, and the current experimental technologies is already prepared for the target. With the ability to realize direct measurement of superposition wavefunction, we get an important opportunity to directly verify the Born rule.

\section{Weak value based direct superposition wavefunction measurement}
The weak value of observable $\hat{A}$ is defined as
\begin{equation}
<\hat{A}>_{w}=\frac{\langle\phi_{f}|\hat{A}|\psi_{i}\rangle}{\langle\phi_{f}|\psi_{i}\rangle},
\end{equation}
where $|\psi_{i}\rangle, |\phi_{f}\rangle$ represent the initial state and post-selection state of the system respectively. If $\hat{A}$ is replaced with position projector $|x\rangle\langle x|$ and $|\phi_{f}\rangle$ be zero momentum state $|p=0\rangle$, then we obtain
\begin{equation}
<|x\rangle\langle x|>_{w}=\frac{\langle p=0|x\rangle\langle x|\psi_{i}\rangle}{\langle p=0|\psi_{i}\rangle}=C\cdot\psi_{i}(x)
\end{equation}
with constant $C=1/\langle p=0|\psi_{i}\rangle$. The real and imaginary part of $<|x\rangle\langle x|>_{w}$ can be extracted directly by performing suitable observables in experiment, which reveals the information of wavefunction $\psi_{i}(x)$.

The extension to the case of superposition wavefunction is direct. Consider $\hat{A}=|x_{1}\rangle\langle x_{1}|+|x_{2}\rangle\langle x_{2}|$, we have
\begin{equation}
<|x_{1}\rangle\langle x_{1}|+|x_{2}\rangle\langle x_{2}|>_{w}=C\cdot[\psi_{i}(x_{1})+\psi_{i}(x_{2})].
\end{equation}
The direct measurement of superposition wavefunction requires direct measurement of superposition weak value $<|x_{1}\rangle\langle x_{1}|+|x_{2}\rangle\langle x_{2}|>_{w}$. We now specify the physical realization of the direct measurement of superposition weak value for more general case. The interaction Hamiltonian between the system and the pointer can be settled as 
\begin{equation}
\hat{H}=\theta\cdot\sum_{i}|x_{i}\rangle\langle x_{i}|\otimes\hat{\sigma}_{y},
\end{equation}
where $\theta=\pi/2$ is the coupling strength and $\hat{\sigma}_{y}$ is the Pauli operator. The unitary evolution corresponding to this Hamiltonian $\hat{U}=e^{-i\hat{H}}$ implies that the qubit pointer be rotated $\pi$ along $Y$ axis of the Bloch sphere if the system in the position $x_{i}$. Suppose that the initial states of the system and pointer are $|\psi\rangle_{s}$ and $|0\rangle_{p}$ respectively, the composite state after the interaction becomes
\begin{equation}
\begin{split}
|\Psi\rangle_{sp} &=\hat{U}|\psi\rangle_{s}\otimes|0\rangle_{p}         \\
&=|\psi\rangle_{s}\otimes|0\rangle_{p}-\sqrt{2}\sum\psi_{s}(x_{i})|x_{i}\rangle\otimes|-\rangle_{p}
\end{split}
\end{equation}
with $|-\rangle=(|0\rangle-|1\rangle)/\sqrt{2}$. The post-selection of the system to the zero momentum state collapse the pointer into the state
\begin{equation}
\begin{split}
|\phi\rangle_{p}&=\langle p=0|\Psi\rangle_{sp}  \\
&=\phi_{s}(p=0)|0\rangle_{p}-\sqrt{2}\sum\psi_{s}(x_{i})\langle p=0|x_{i}\rangle\otimes|-\rangle_{p}  \\
&=\frac{1}{C}[|0\rangle_{p}-\sqrt{2}<\sum|x_{i}\rangle\langle x_{i}|>_{w}|-\rangle_{P}].
\end{split}
\end{equation}
The real and imaginary part of superposition weak value $<\sum|x_{i}\rangle\langle x_{i}|>_{w}$ can be directly extracted by the proper measurements. Specifically, we have
\begin{equation}
\begin{split}
&\mathrm{Re}[<\sum|x_{i}\rangle\langle x_{i}|>_{w}]=\frac{1}{2|C|^{2}}\langle\phi|(\hat{\sigma}_{x}+2|1\rangle\langle 1|)|\phi\rangle_{p}        \\
&\mathrm{Im}[<\sum|x_{i}\rangle\langle x_{i}|>_{w}]=\frac{1}{2|C|^{2}}\langle\phi|\hat{\sigma}_{y}|\phi\rangle_{p}.
\end{split}
\end{equation}
The above process for superposition wavefunction reconstruction is totally same with the previous wavefunction reconstruction \cite{wave1},  but only with $|x\rangle\langle x|$ be replaced with $\sum|x_{i}\rangle\langle x_{i}|$. 

\begin{figure}[tbp]
\centering
\includegraphics[scale=0.75]{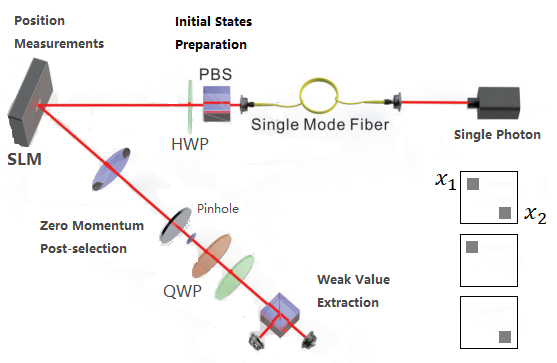}
\caption{Experimental setup proposal for direct measurement pf superposition wavefunction and thus directly verification of the Born rule. The single photons are prepared in the polarization state $|0\rangle_{p}=(|H\rangle+|V\rangle)/\sqrt{2}$ initially, and then reflected by a phase-type SLM in which $\pi$ phase difference between $|H\rangle$ and $|V\rangle$ can be added in the chosen areas. Photons that reflected from the chosen areas are converted into the orthogonal polarization state $|1\rangle_{p}=(|H\rangle-|V\rangle)/\sqrt{2}$, which realizes the position $|x\rangle\langle x|$ measurements. The post-selection of zero momentum state is completed by a Fourier lens and a pinhole. Finally the weak value is directly extracted by measuring the proper observables using the polarization analyser. The superposition wavefunction $\psi(x_{1})+\psi(x_{2})$ is obtained by adding phases in areas $x_{1}, x_{2}$ simultaneously, while $\psi(x_{1}), \psi(x_{2})$ are obtained by adding phase only in $x_{1}$ and $x_{2}$ respectively. {\bf PBS}: polarization beam splitter, {\bf HWP}: half wave plate, {\bf QWP}: quarter wave plate, {\bf SLM}: spatial light modulator.  }
\label{f2}
\end{figure}

\section{Experimental proposal}
The experimental setup to directly measure the superposition wavefunction and thus verify the Born rule is almost the same as the one reported in our previous work \cite{wave1}. There are, however, two differences. First, there is no need to prepare specific spatial wavefunction of photons. Any state of the system $|\psi\rangle_{s}$ is acceptable given that the preparation is the same during all the runs. Second, there is no need to scan the position, and we just add the $\pi$ phase in chosen area $x_{1}$ or $x_{2}$ or simultaneously in sequence.

The setup, as shown in Fig. 1, consists of four parts, i.e., preparation of initial states, measurement of position, post-selection of zero-momentum state and extraction of weak value. Single photons can be generated by the spontaneous parametric down-conversion process with the trigger \cite{Zhu}. The pointer is chosen as the polarization of photons, which is prepared in the state $|0\rangle_{p}=(|H\rangle+|V\rangle)/\sqrt{2}$ by passing through a PBS and a HWP rotated at $22.5^{\circ}$. The photons are reflected by a phase-only SLM in which $\pi$ phase is added in the chosen areas. The polarization state of photons that reflected from these areas will be converted into orthogonal state $|1\rangle_{p}=(|H\rangle-|V\rangle)/\sqrt{2}$. This completes the position $|x\rangle\langle x|$ measurement because we can know definitely whether or not photons have passed the position $x$ according to the polarization state of photons. The post-selection of zero momentum state is realized by a Fourier lens and a pinhole placed in its focal plane. The weak value is extracted by measuring the polarization pointer with a polarization analyser consists of HWP, QWP and PBS. The direct quantites that we obtained in experiment is the expectation values $\langle\phi|\hat{\sigma}_{x}|\phi\rangle_{p}, \langle\phi|\hat{\sigma}_{y}|\phi\rangle_{p}$ and $\langle\phi|1\rangle\langle 1|\phi\rangle_{p}$. It should be noted that $\hat{\sigma}_{x}, \hat{\sigma}_{y}$ are defined in the basis of $\lbrace|0\rangle, |1\rangle\rbrace$.
Since weak value is proportional to the wavefunction and the constant $C$ is fixed in experiment for all runs, we could omit the constant and write the superposition wavefunction with the observed expectation values as  
\begin{equation}
\begin{split}
&\mathrm{Re}[\sum\psi(x_{i})]= \langle\phi|(\hat{\sigma}_{x}+2|1\rangle\langle 1|)|\phi\rangle_{p}    \\
&\mathrm{Im}[\sum\psi(x_{i})]= \langle\phi|\hat{\sigma}_{y}|\phi\rangle_{p}.
\end{split}
\end{equation}

\section{Discussion and Conclusion}
Currently, there seems no technical difficulty to complete the suggested experiment. However, it would not be easy to obtain accurate results. 
The key is to reduce the system error introduced by the imperfect instruments as small as possible. The statistical error can be reduced significantly by the cumulative counting. The accurate experimental results will tells us to what extent the Born rule applies.

In conclusion, we have introduced the weak value based direct measurement of superposition wavefunction and its potential to the direct verification of the Born rule. The specific experimental setup of photonic system to fulfill the target is proposed. 

\hfill

\begin{acknowledgments}
Meng-Jun Hu acknowledges the fruitful discussions with Prof. Yong-Sheng Zhang. Meng-Jun Hu especially thanks the supported from the Bejing Academy of Quantum information Sciences.

\end{acknowledgments}


\begin{thebibliography}{0}%
\makeatletter
\providecommand \@ifxundefined [1]{%
 \@ifx{#1\undefined}
}%
\providecommand \@ifnum [1]{%
 \ifnum #1\expandafter \@firstoftwo
 \else \expandafter \@secondoftwo
 \fi
}%
\providecommand \@ifx [1]{%
 \ifx #1\expandafter \@firstoftwo
 \else \expandafter \@secondoftwo
 \fi
}%
\providecommand \natexlab [1]{#1}%
\providecommand \enquote  [1]{``#1''}%
\providecommand \bibnamefont  [1]{#1}%
\providecommand \bibfnamefont [1]{#1}%
\providecommand \citenamefont [1]{#1}%
\providecommand \href@noop [0]{\@secondoftwo}%
\providecommand \href [0]{\begingroup \@sanitize@url \@href}%
\providecommand \@href[1]{\@@startlink{#1}\@@href}%
\providecommand \@@href[1]{\endgroup#1\@@endlink}%
\providecommand \@sanitize@url [0]{\catcode `\\12\catcode `\$12\catcode
  `\&12\catcode `\#12\catcode `\^12\catcode `\_12\catcode `\%12\relax}%
\providecommand \@@startlink[1]{}%
\providecommand \@@endlink[0]{}%
\providecommand \url  [0]{\begingroup\@sanitize@url \@url }%
\providecommand \@url [1]{\endgroup\@href {#1}{\urlprefix }}%
\providecommand \urlprefix  [0]{URL }%
\providecommand \Eprint [0]{\href }%
\providecommand \doibase [0]{http://dx.doi.org/}%
\providecommand \selectlanguage [0]{\@gobble}%
\providecommand \bibinfo  [0]{\@secondoftwo}%
\providecommand \bibfield  [0]{\@secondoftwo}%
\providecommand \translation [1]{[#1]}%
\providecommand \BibitemOpen [0]{}%
\providecommand \bibitemStop [0]{}%
\providecommand \bibitemNoStop [0]{.\EOS\space}%
\providecommand \EOS [0]{\spacefactor3000\relax}%
\providecommand \BibitemShut  [1]{\csname bibitem#1\endcsname}%
\let\auto@bib@innerbib\@empty
\end{thebibliography}%


\begin{thebibliography}{99}

\bibitem{AAV} Y. Aharonov, D. Z. Albert, and L. Vaidman, How the result of a measurement of a component of the spin of a spin-1/2 particle can turn out to be 100, Phys. Rev. Lett {\bf 60}, 1351 (1988).

\bibitem{Lundeen} J. S. Lundeen, B. Sutherland, A. Patel, C, Stewart, and C. Bamber, Direct measurement of the quantum wavefunction, Nature {\bf 474}, 188-191 (2011).

\bibitem{wave1} C. R. Zhang, M. J. Hu, Z. B. Hou, J. F. Tang, J. Zhu, G. Y. Xiang, C. F. Li, G. C. Guo, and Y. S. Zhang, Direct measurement of the two-dimensional spatial quantum wave function via strong measurements, Phys. Rev. A {\bf 101}, 012119 (2020).

\bibitem{wave2} Z. M. Shi, M. Mirhosseini, J. Margiewicz, M. Malik, F. Rivera, Z. Y. Zhu, and R. W. Boyd, Scan-free direct measurement of an extremely high-dimensional photonic state, Optica {\bf 2}, 4 (2015).

\bibitem{wave3} C. R. Zhang, M. J. Hu, G. Y. Xiang, Y. S. ZHang, C. F. Li, and G. C. Guo, Direct Strong Measurement of a High-Dimensional Quantum State, Chin. Phys. Lett. {\bf 37}, 080301 (2020).

\bibitem{wave4} Y. Zhou, J. Zhao, D. Hay, K. McGonagle, R. W. Boyd, and Z. Shi, Direct Tomography of High-Dimensional Density Matrices for General Quantum States of Photons, Phys. Rev. Lett. {\bf 127}, 040402 (2021).

\bibitem{wave5} G. Vallone and D. Dequal, Strong Measurements Give a Better Direct Measurement of the Quantum Wave Function, Phys. Rev. Lett. {\bf 116}, 040502 (2016).

\bibitem{Zhu} J. Zhu, M. J. Hu, S. M. Cheng, M. J. W. Hall, C. F. Li, G. C. Guo, and Y. S. Zhang, Experimental verification of anisotropic invariance for three-qubit states, Phys. Rev. A {\bf 99}, 040103(R) (2019).


\end{thebibliography}
\end{document}